\documentclass{jfm}
\usepackage[utf8]{inputenc}
\usepackage[breaklinks]{hyperref}
\usepackage{graphicx}
\usepackage{natbib}
\usepackage{outlines}
\usepackage{hyperref}

\usepackage{indentfirst}
\usepackage{amsmath}
\usepackage{xfrac}
\usepackage{subfigure}

\usepackage[usenames,dvipsnames]{pstricks}
\usepackage{epsfig}
\usepackage{pst-grad} % For gradients
\usepackage{pst-plot} % For axes

\ifCUPmtlplainloaded \else
  \checkfont{eurm10}
  \iffontfound
    \IfFileExists{upmath.sty}
      {\typeout{^^JFound AMS Euler Roman fonts on the system,
                   using the 'upmath' package.^^J}%
       \usepackage{upmath}}
      {\typeout{^^JFound AMS Euler Roman fonts on the system, but you
                   dont seem to have the}%
       \typeout{'upmath' package installed. JFM.cls can take advantage
                 of these fonts,^^Jif you use 'upmath' package.^^J}%
      }
  \else
  \fi
\fi

\ifCUPmtlplainloaded \else
  \checkfont{msam10}
  \iffontfound
    \IfFileExists{amssymb.sty}
      {\typeout{^^JFound AMS Symbol fonts on the system, using the
                'amssymb' package.^^J}%
       \usepackage{amssymb}%

      }{}
  \fi
\fi

\ifCUPmtlplainloaded \else
  \IfFileExists{amsbsy.sty}
    {\typeout{^^JFound the 'amsbsy' package on the system, using it.^^J}%
     \usepackage{amsbsy}}
    {}
\fi

\newsavebox{\astrutbox}
\sbox{\astrutbox}{\rule[-5pt]{0pt}{20pt}}

\usepackage{ulem}
\def\ADD#1{{\textcolor{black}{#1}}}    % suggested new text
\title[]{Internal wave attractors examined using laboratory experiments and 3D numerical simulations.}

\author{C.~Brouzet$^1$, I.~Sibgatullin$^{1,2,3}$, H. Scolan$^{1,4}$, E.V.~Ermanyuk$^{1,5}$, T.~Dauxois$^{1}$}

\affiliation{1. Laboratoire de Physique, ENS de Lyon, 
Universit\'e de Lyon, CNRS, 
Lyon, France,\\
2. Institute of Mechanics and Department of Mechanics and Mathematics, Moscow State University, Russia\\
3. Institute for System Programming, Moscow, Russia
\\
4. Atmospheric, Oceanic and Planetary Physics, Department of Physics, University of Oxford, Oxford, UK.\\
5. Lavrentyev Institute of Hydrodynamics, Novosibirsk, Russia\\
}

\pubyear{2016}
\begin{document}

\maketitle

\begin{abstract}
In the present paper, we combine numerical and experimental approaches to study the dynamics of stable and unstable internal wave attractors. The problem is considered in a classic trapezoidal setup filled with a uniformly stratified fluid. Energy is injected into the system at global scale by the small-amplitude motion of a vertical wall. Wave motion in the test tank is measured with the help of conventional synthetic schlieren and PIV techniques. The numerical setup closely reproduces the experimental one in terms of geometry and the operational range of the Reynolds and Schmidt numbers. The spectral element method is used as a numerical tool to simulate the nonlinear dynamics of a viscous  salt-stratified fluid. We show that the results of three-dimensional calculations are in excellent qualitative and quantitative agreement with the experimental data, including the spatial and temporal parameters of the secondary waves produced by triadic resonance instability. 
Further, we explore experimentally and numerically the effect of lateral walls on secondary currents and spanwise distribution of velocity amplitudes in the wave beams. Finally, we test the assumption of a bidimensional flow and estimate the error made in synthetic schlieren measurements due to this assumption.
\end{abstract}

\section{Introduction}
Stratified and/or rotating fluids are ubiquitous in geophysical and astrophysical hydrodynamics. Such fluids support internal and/or inertial waves which have many common properties due to the similarity of their dispersion relations: typical occurrences of these waves are oblique beams that propagate at an angle with respect to the horizontal controlled by the ratio of the forcing frequency to the natural frequency of the system, the latter being set by the combined balances  of restoring (buoyancy and/or Coriolis) force and inertia.

 The anisotropic dispersion relation requires the preservation of the angle of the wave beam to the horizontal upon reflection at a rigid boundary. In the case of a sloping boundary, this property gives a geometric reason for the strong variation of the beam width (focussing or defocussing) upon reflection. In the limit of a vanishingly small beam width, the wave reflection in confined domains can be studied by ray tracing which shows that focussing prevails. A typical generic case is represented by  wave rays converging to a closed loop, the wave attractor. A detailed analysis shows that a confined domain of a specific geometry has a subset of global resonances (similar to normal modes) and a rich variety of wave attractors having different complexity and rates of convergence toward limit cycles~\citep{MaasLam1995}.

Internal wave attractors have been experimentally observed in the seminal paper by~\cite{MBSL1997} for a benchmark geometric setting, a trapezoidal domain filled with a uniformly stratified fluid. The properties of internal wave attractors have been extensively studied~\citep{GSP2008,HBDM2008,HTMD2008,EYBP2011,HGD2011}. In parallel, a rich literature describes the properties of  inertial wave attractors, considering a rotating spherical annulus as a generic model of homogeneous liquid shells of celestial bodies~\citep{Stewartson1971,Stewartson1972,RieutordValdettaro1997,RGV2000,RGV2001,RVG2002,RieutordValdettaro2010,RabitiMaas2013}, and also simpler geometrical settings of academic interest~\citep{MandersMaas2003,Ogilvie2005,JouveOgilvie2014}. Inertia-gravity waves in rotating stratified fluids contained in spherical shells have also been considered~\citep{DRV1999,MaasHarlander2007}. In addition to the above-mentioned two-dimensional (plane or axisymmetric) settings, fully three-dimensional geometries have been considered theoretically  ~\citep{Maas2005},  numerically \citep{DM2007} and experimentally ~\citep{HazewinkelMaasDalziel2011,MandersMaas2004}.

The major part of the cited literature is focussed on the theoretical analysis of wave attractors, assuming linearized equations of motion and a simple harmonic time-dependence. In the inviscid limit, the theory explores the structure of singular and regular solutions of the wave equation (in spatial coordinates) in a confined domain, what is known to be generally an ill-posed problem: see e.g.~\cite{Stewartson1971,Stewartson1972,MaasLam1995,RieutordValdettaro1997}. The inclusion of viscosity regularizes the problem, replacing attractors singularities by shear layers of finite width, which can be computed numerically~\citep{RieutordValdettaro1997,RGV2001}. The competition between geometric focussing and diffusive viscous broadening provides a physically clear mechanism for setting the final width of the branches of wave attractors, at least in linear problems~\citep{Ogilvie2005,HBDM2008,GSP2008}. Other dissipative mechanisms can be involved, which might be modeled by considering a relevant eddy viscosity, 
 triadic resonance instability accompanied with energy flux to waves with shorter length scales, or magnetic damping in the case of conductive fluids~\citep{Ogilvie2005}. Remaining within the realm of the linearized approach, \cite{Ogilvie2005} has shown that the asymptotic dissipation rate in the system (i.e. the value corresponding to the zero viscosity limit or, equivalently, a vanishingly small Ekman number) is, to the leading order, not sensitive to the replacement of viscous damping by a 'frictional' damping force proportional to the velocity. However, if a system is given a sufficiently large forcing, nonlinear effects become important, motivating experiments and direct numerical simulations of the fully nonlinear problem.

The first {numerical simulations of the corresponding Navier-Stokes equation} have been performed in~\cite{GSP2008} and a subsequent study has been presented in~\cite{HGD2011}. It has shown that numerical results successfully reproduce the experimentally observed dynamics of wave beam formation~\citep{HBDM2008} in terms of the overall flow pattern, width of the wave beams and  shape of the wave number spectra during transient and steady regimes.
\cite{HGD2011} have reported also a good qualitative agreement in terms of measured and computed stream functions, and have concluded that the good quantitative agreement for stable attractors is possible if the forcing in the numerical model is adjusted in such a way that the computed peak velocities coincide with the measured ones (it relies on a forcing in 2D numerical simulation much lower than in the experiments.). Regarding the computed non-linear effects, \cite{GSP2008} have described the harmonic components corresponding to multiples (double and triple) of the forcing frequency. The presence of the second harmonic has also been experimentally detected in the data from~\cite{MBSL1997}, and later re-analyzed in~\cite{LamMaas2008}.

The stability of attractors 
for an injected energy of high amplitude 
 remained largely unexplored until the experimental work by~\cite{SED2013}, which has demonstrated that attractors are prone to triadic resonance instability Note that the particular case for which both unstable secondary waves have a frequency equal to half of the forcing frequency is of particular interest in the oceanographic context where viscosity is negligible. In that case, the appropriate name is parametric subharmonic instability and abbreviated as PSI.
By abuse of language, some authors have sometimes extended the use of the name PSI to cases for which secondary waves are not corresponding to half of the forcing frequency. For the sake of terminological consistency, we propose to abbreviate triadic resonance instability in the rest of the paper using the new acronym: TRI.
 
 The onset of the instability is similar to the scenario for wave beams in an unbounded domain~\citep{BDJO2013}: 
 in reality 
  the finiteness of the beam width matters~\citep{KarimiAkylas2014,BSDBOJ2014}. Direct numerical two-dimensional simulations of inertial wave attractors in linear and nonlinear regimes have been performed in~\cite{JouveOgilvie2014} for a model geometry represented by a tilted square. In the linear regime, the simulations reported very nicely confirm the linear analysis presented in~\cite{Ogilvie2005}. In the nonlinear regime, the instability is shown to transfer energy to short wavelengths where it is more efficiently dissipated by viscosity. Owing to this mechanism, the saturation of the overall dissipation in the system occurs faster than in the linear regime. The scenario of instability observed in these numerical simulations~\citep{JouveOgilvie2014} for inertial waves is qualitatively similar to the one described in~\cite{SED2013} for internal waves. In the oceanographic context, observations have provided strong suggestions 
of the particular case of parametric subharmonic instability ~\citep{MacKinnonWinters2005,AMZPKP2007,Alford2008}.

Summing up the literature survey, we note that there are only few studies reporting direct numerical simulations of internal~(\cite{GSP2008};
\cite{HGD2011}) and inertial~\citep{JouveOgilvie2014} wave attractors. Except in \cite{DM2007} in a more oceanographic context, all these studies were performed in two-dimensional settings.
\cite{GSP2008}  use a trapezoidal geometry, with one horizontal boundary representing a free surface, and free-slip
conditions at rigid boundaries of the fluid volume, except at the vertical wall, where barotropic forcing is applied in the form of a uniform
horizontal flow of small amplitude oscillating at a chosen forcing frequency. The implementation of free-slip conditions  
avoid resolving viscous boundary layers. 
\cite{HGD2011} use a similar geometrical setup but forced the system via a progressive first-mode internal wave.
In both cases, \cite{GSP2008} and~\cite{HGD2011} take advantage of the two-dimensional version of MITgcm, the general circulation model developed by~\cite{MAHPH1997} based on finite volume method.
 The Prandtl-Schmidt
number (the ratio of the fluid viscosity to the salt diffusivity) was taken equal to 100 in~\cite{GSP2008} and 770 in~\cite{HGD2011}. In their study of inertial waves,
\cite{JouveOgilvie2014} use a different geometry: a tilted square. 
They carried out direct numerical simulations using a 
 two-dimensional version of the 3D spectral code
SNOOPY~\citep{LesurLongaretti2005,LesurLongaretti2007,LesurOgilvie2010}. No-slip boundary conditions are 
 imposed 
 via a fictitious
absorbing layer outside the fluid domain where the velocity components are forced to vanish, ensuring the global energy conservation with an
accuracy of a few per cent.

To our knowledge, there have been no attempts of 3D simulations 
for direct comparison with experimentally generated attractors.
In rectangular domains where normal modes are of primary interest, the dissipation at rigid walls (boundary friction) can dominate the dissipation in the bulk by a factor of one hundred \citep{BenielliSommeria1998,LamMaas2008}. The importance of boundary friction for triadic resonance instability 
 in the case of normal modes in a rectangular domain has been already emphasized in~\cite{McEwan1971}. 
  His experimental results of the amplitude threshold for TRI of normal modes are indeed in good agreement with theoretical predictions when the sum of a bulk-internal term and a boundary-layer term is taken into account in the calculated energy dissipation. Moreover, he noticed that, on a laboratory scale, internal dissipation becomes comparable with wall dissipation only from a vertical modal number of order 10. 
 In the case of the normal modes, an analytical expression for the energy dissipation can be 
  found and reveals that  the bulk dissipation depends on the wavelength of the wave while the boundary dissipation is only related to the direction of the wavenumber vector.
   In contrast to normal modes, attractors are strongly dissipative structures which adapt their typical wavelength (and, therefore, shear and associated dissipation rate)  to reach a global balance between the injected and the dissipated energy~\citep{Ogilvie2005}. Thus, the role of the dissipation at rigid boundaries as compared to the dissipation by shearing in the bulk of fluid is less clear for attractors than for normal modes. 
   It is also not obvious to what extent 3D simulations are necessary to reproduce the experimental results quantitatively. 
   The present paper aims at addressing these questions about dissipation and 3D effects by performing 
    cross-comparisons of the available experimental data against the results of three-dimensional direct numerical simulations of internal wave attractors using  a spectral element method based on the code Nek5000~\citep{FischerRonquist1994}.

The paper is organized as follows. In Section 2, we describe the experimental and numerical setup. In Section 3, we compare the results of three-dimensional computations with the experimental data in the linear and nonlinear regimes. In particular, in the nonlinear regime, we apply to numerical results the technique previously employed in~\cite{SED2013} for analyzing triadic resonances. In Section 4, we describe the wave structure in the transversal direction. Using experiments and numerical simulations, we test the assumption of the bidimensionality of the flow and we examine the mean-flow generated by the wave attractor. 
We also  estimate the error made in synthetic schlieren measurements due to the bidimensional flow assumption.
Finally,  in section 5, we present our conclusions.

\section{Experimental and numerical set-ups}

\subsection{Experimental set-up}

\begin{figure}
% \begin{center} \includegraphics[width=0.7\linewidth,clip=]{setup_droit.eps}\end{center}
\begin{center} \includegraphics[width=0.7\linewidth,clip=]{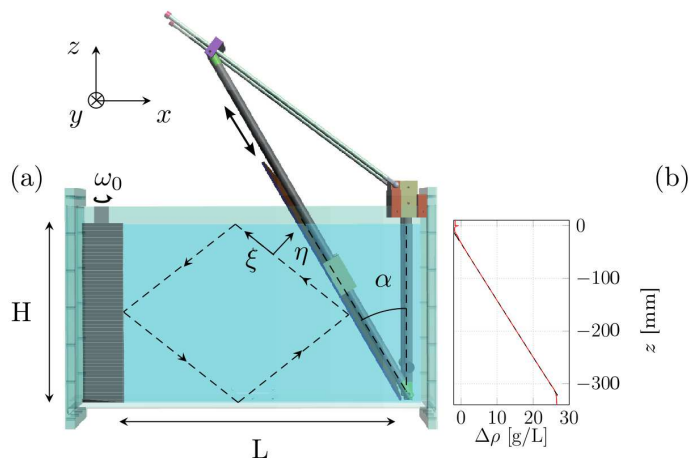}\end{center}
  \vspace{-.3cm}
   \caption{(a): Experimental set-up.  A sloping wall is inserted inside
a tank of size L$\times$W$\times$H~=~800$\times$170$\times$425 \ADD{mm}$^3$. The working bottom
length~$L$ of the section, the depth~$H$, and the sloping angle~$\alpha$
can be clearly identified in this picture.  A wave maker located on the left of the tank
generates a mode~1 forcing at a tunable frequency~$\omega_0$. The flow is mostly
two-dimensional as demonstrated
 within the text and, therefore, essentially independent of the transversal variable~$y$
 except for narrow boundary layers at lateral walls and weak nearly horizontal secondary currents in the wave beams.
The schematic attractor is depicted with the dotted quadrilateral in the working domain.
The longitudinal (resp. transversal) variable $\xi$ (resp. $\eta$) of the most energetic branch of the attractor
are also defined on the picture. (b): Example of a stratification measured with the conductivity probe before one experiment. The density difference with fresh water, $\Delta \rho$, is plotted in function of the depth $z$.
 \label{ExperimentalSetup}
   }
\end{figure}

In the present work, we use the experimental results presented in~\cite{SED2013} and also additional experimental data obtained with the experimental setup sketched in figure \ref{ExperimentalSetup}(a). A cartesian coordinate system is introduced, with the horizontal $x$ and vertical $z$ axes located in the vertical midplane of the test tank. The $y$ axis is perpendicular to the $(x,z)$-plane and directed from the observer/camera towards the tank. The rectangular test tank of size $800\times 170\times 425 ~$\ADD{mm}$^{3}$ is filled with a salt-stratified fluid using the conventional double-bucket technique.
The density distribution as a function of the vertical coordinate $z$ is measured prior and after experiments by a conductivity probe driven by a vertical traverse mechanism. The value of the buoyancy frequency $N=[(-g/\bar{\rho})({\rm d}\rho/{\rm d}z)]^{1/2}$ is inferred from the linear fit to the measured density profile. An example of such a profile is shown on figure~\ref{ExperimentalSetup}(b). A sliding sloping wall, inclined at the angle $\alpha$,  is carefully inserted into the fluid once the filling procedure is over. The sloping wall delimits a trapezoidal fluid domain of length~$L$ (measured along the bottom) and depth $H$.

The input forcing is introduced into the system by an internal wave generator~\citep{GDMD2007,MMMGPD2010,JMOD2012,SED2013} with the time-dependent vertical profile given by
\begin{equation}
\zeta(z,t)=a\sin(\omega_{0} t) \cos(\pi z/H),
 \label{generator-profile}
\end{equation}
where $a$ and $\omega_{0}$ are the amplitude and frequency of oscillations, respectively. The profile
is reproduced in discrete form by the motion of a stack of $47$ horizontal plates driven by the rotation of a vertical shaft. The stack of moving plates is confined between lateral walls of a vertical box-like support frame. Each lateral wall of the frame is $15$~mm thick. Therefore the inner size of the frame, $W'$, which coincides with the width of the moving plates, is $30$~mm smaller than the width of the tank $W$. Thus the forcing is applied to the stratified fluid with a width $W'$ representing $82\%$  of the whole width $W$.

Fluid motion is recorded by a computer-controlled video  AVT (Allied Vision Technologies) Stingray camera  with CCD matrix of $1388 \times 1038$ pixels or Pike camera with CCD matrix of $2452 \times 2054$. The camera is placed at a distance between 1750 to  $2300$~\ADD{mm} from the test tank. Given the relatively small size of the working section, no correction for the parallax errors is applied.
Two methods are used to measure the parameters of the fluid motion. The perturbations of the density gradient are measured with 
synthetic schlieren technique~\citep{DHS2000} from apparent distortions of a background random dot pattern observed through the stratified fluid.  In some cases,
standard PIV technique  has alternatively been used for velocity measurements. The fluid is seeded with light-reflecting particles
of typical size $8~\mu$m and density $1.1$~kg/m$^{3}$. The sedimentation velocity of particles is found to be sufficiently low. Its effect on the results of velocity measurements is negligibly small. The test section is illuminated by a vertical laser sheet coming through the transparent side or bottom of the test tank. The spanwise position of the sheet could be varied to assess three-dimensional effects.

For the synthetic schlieren technique, the camera takes $2$ frames per second.
For the PIV measurements, the camera operates in two-frame burst mode with an interval between the pair of images within a burst set to $0.125$~s and a global burst cycle period of $0.5$~s. Conversion of images into velocity fields is performed using a cross-correlation PIV algorithm with subpixel resolution~\citep{FinchamDelerce2000}. Thus with $2$ velocity fields per second, i.e typically around $20$ fields per wave period, we can resolve the significant frequency components of the signal. Besides, the spatial resolution for the velocity field is about 3 mm in each direction and is found sufficient to resolve fine details of the wave field.

\subsection{Numerical set-up}

The numerical simulation of wave attractors faces two major challenges. First, the fluid motion is highly nonlinear, and accounting for nonlinear interactions is crucial for the dynamics, even for weak interactions. Second, the Schmidt number ($Sc$), defined as the ratio of water kinematic viscosity and salt diffusivity, is close to 700 in a salt-stratified fluid. So the spatial scale of the density perturbations can be much smaller than the scale of the velocity perturbations 
and more demanding in terms of spatial resolution.
{The scale under which no scalar gradient remains because of diffusion effect is very small and smaller than the usual Kolmogorov scale $L_K$ used as a mesh criterion for direct simulations where $L_K = Re^{-3/4} \times L$ with $L$ the integral scale and Re the large scale Reynolds number. Strictly speaking, this small scale where scalar diffusion takes place can be estimated with the Batchelor scale $L_B = L_K \times  Sc^{-1/2}$ ~\citep{Batchelor1959,BuchDahm1996,Rahmani2014}.}

The first numerical simulation of the formation of internal wave attractors~\citep{GSP2008} used finite volume method, one of the most popular state-of-the-art tools in computational fluid dynamics. However, as noted by these authors, the numerical simulation could only reproduce the dynamics of the attractors for Schmidt number less than 100. In addition, the discretization of convective terms produces a numerical viscosity, which acts similarly to a real viscosity. This numerical artefact blurs the fine-scale structures arising due to the high value of the Schmidt number and it may also introduce substantial errors in calculations of the long-term dynamics of attractors (at time scales of tens or hundreds of periods).
However, {for the current study,} the long time intervals are precisely of particular interest to study the growth, saturation and cascades of secondary instabilities. 

As an alternative, the Galerkin method represents a highly efficient approach to tackle the nonlinear interactions if a full system of basis functions subject to boundary conditions can be used. A pseudospectral approach offers an efficient realization of the Galerkin method. In \cite{JouveOgilvie2014}, this approach has been used to study  nonlinear interactions in attractors of inertial waves. In the present work we have chosen the method of spectral elements which combines the accuracy and high resolution of spectral methods with geometric flexibility of finite element methods, and what is particularly suitable for simulation of long-term evolution of fine-scale flows in globally forced geophysical systems~\citep{Favier-etal-2015}. The computational domain is divided into a finite number of quadrilateral (in 2D) or hexahedral (in 3D) elements and a Galerkin approach is applied to each element. For the numerical realization, we use the open code Nek5000, developed by Paul Fischer and colleagues~\citep{FischerRonquist1994,Fischer1997,FischerMullen2001}. In each element, the Lagrange polynomial decomposition is used and applied at Gauss-Lobatto-Legendre points for the sake of stability (to avoid ill-conditioning). The full resulting mesh, consisting of the elements and the Gauss-Lobatto-Legendre points, is highly nonuniform, adding a bit more complexity in post-processing data treatment.
However, the efficiency of the code fully justifies such a nuisance.  In other words, the approach used weighted residual techniques employing tensor-product polynomial bases.
Besides other benefits, it allows ``analytical" computation of the derivatives through matrix-matrix products or matrix-matrix-based evaluation
\begin{equation}
  u(x,y,z,t)=\sum_{i,j,k} u_{i,j,k}(t) \psi_i(x) \psi_j(y) \psi_k(z)
\end{equation}
\begin{equation}
  \frac{\partial}{\partial x}u(x,y,z,t)=\sum_{i,j,k} u_{i,j,k}(t) \psi_i^{'}(x) \psi_j(y) \psi_k(z),
\end{equation}
in which $u$ represents any of the unknown variables, for instance, velocity components, density, etc. $\psi$ is a Lagrangian interpolant through the
Gauss-Lobatto-Legendre points.

The implementation of a high-order filter in Nek5000 allows to stabilize the method for convection-dominated  flows~\citep{FischerMullen2001}.  The time advancement is based on second-order semi-implicit operator splitting methods and stable backward-difference scheme. The additive overlapping Schwarz method is used as a pre-conditioner~\citep{Fischer1997}.

The most difficult regions from a computational viewpoint are located in the vicinity of the rigid walls since, in these regions, intense folding of high-gradient density layers may occur, especially in the case of a high-amplitude wave motion. For this reason, a nonuniform element mesh in the near-wall regions is preferable for simulations of the nonlinear dynamics of wave attractors. Typically, we have used meshes with up to 0.5 million elements, with eighth to tenth order polynomial decomposition within each element.

The full system of equations being solved consists of the Navier-Stokes equation in the Boussinesq approximation
\begin{equation}
\rho_m \left(\frac{\partial 
{\mbox{\boldmath $v$}}}{\partial t} + (\mbox{\boldmath $v$}\cdot \mbox{\boldmath $\nabla$})  
{\mbox{\boldmath $v$}} \right)= -{\mbox{\boldmath $\nabla$}} {\tilde p} + \mu \Delta 
{\mbox{\boldmath $v$}} + \rho_s  {\mbox{\boldmath $g$}},  
  \label{NSdim}
\end{equation}
the continuity equation
\begin{equation}
 \mbox{\boldmath $\nabla$}\cdot \, 
 \mbox{\boldmath $v$} = 0
  \label{diffDimpp}
\end{equation}
 and the equation for the transport of salt
\begin{equation}
 \frac{\partial \rho_s}{\partial t} + (\mbox{\boldmath $v$}\cdot \mbox{\boldmath $\nabla$}) \rho_s = \lambda \Delta \rho_s,
  \label{diffDim}
\end{equation}
where $\rho_m$ is the density of solution with the constant minimal reference salinity, $\rho_s$  the density perturbation in a unit volume due to local salinity (the full density is $\rho=\rho_m+\rho_s$), $\mu$ the dynamical viscosity and  $\lambda$ the diffusivity of salt. Dynamical viscosity and diffusivity are assumed to be constant.
We impose no-slip boundary conditions on the rigid surfaces and stress-free condition on the upper surface. %}
The boundary conditions on $\rho_s$ are isolation:
$\partial \rho_s / \partial n = 0$
where $n$ is the normal to the wall.
 Forcing is applied at the vertical wall by prescribing the profile of the horizontal velocity which reproduces the motion of the generator~(\ref{generator-profile}) in the vertical direction and takes into account the difference between the width of the generator $W'$ and the width of the tank $W$. The transverse profile of the forcing is prescribed by stepwise or piece-wise linear function. At the central segment of the width $W'$, the horizontal velocity is uniformly distributed in the transverse direction in both cases. At the side segments of width $(W-W')/2$, we prescribe either i) zero velocity or ii) a linear decrease of the velocity from the uniform value to zero. 
 These two versions of transverse profiles of forcing 
  give essentially the same results, with a small quantitative difference in the second version where forcing gives a higher horizontal impulse to the system. Both versions can be implemented 
  in calculations, but the second version of the forcing is found to be much less expensive computationally and the major part of computations was performed with it. 
The efficiency of the generator used in experiments is not 100\%. For this reason, the amplitude of forcing in computations with the piecewise linear case is intentionally reduced by 10\% compared to the experimental value, thereby providing a good match between numerical and experimental results.

The comparison between numerical and experimental results is typically complicated by the presence of a thin mixed layer close to the free surface of the stratified fluid in the test tank. 
Because of this layer, the internal wave beam is not reflected precisely at the free surface. Instead the wave beam undergoes a complex reflection, partially at the interface between the mixed layer and the linearly stratified fluid and partially at the free surface. This complex reflection affects the  shape of the attractor~\citep{GSP2008} and the shape of envelopes of wave motions in the wave beams. 
To take this layer into account
 in the numerical model, we introduce a model density distribution where the full depth is denoted $H$ as in experiments, the thickness of the thin mixed upper layer is $\delta$, and the depth of the linearly stratified fluid is $H'=H-\delta$.

\section{Wave pattern in the vertical plane}

\subsection{A stable attractor}

\begin{figure}
  \includegraphics[width=\linewidth,clip=]{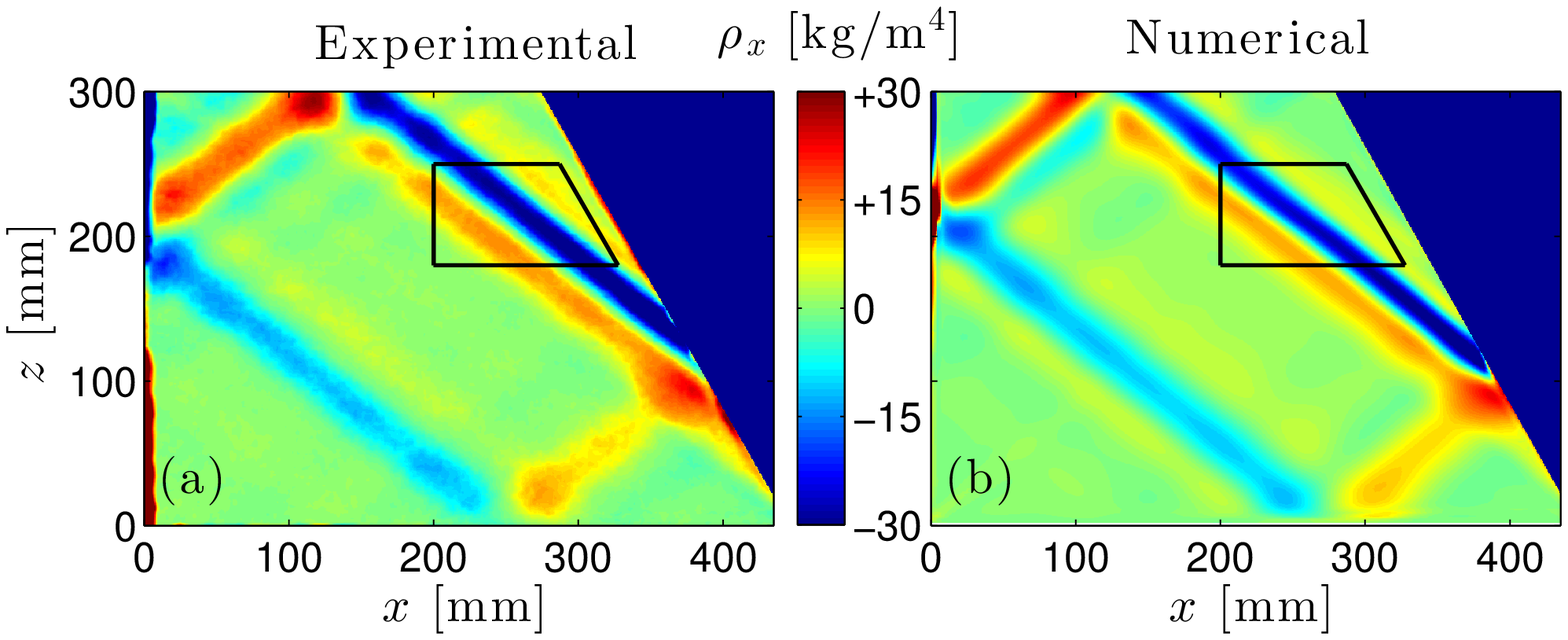}
  \vspace{-.3cm}
  \includegraphics[width=\linewidth,clip=]{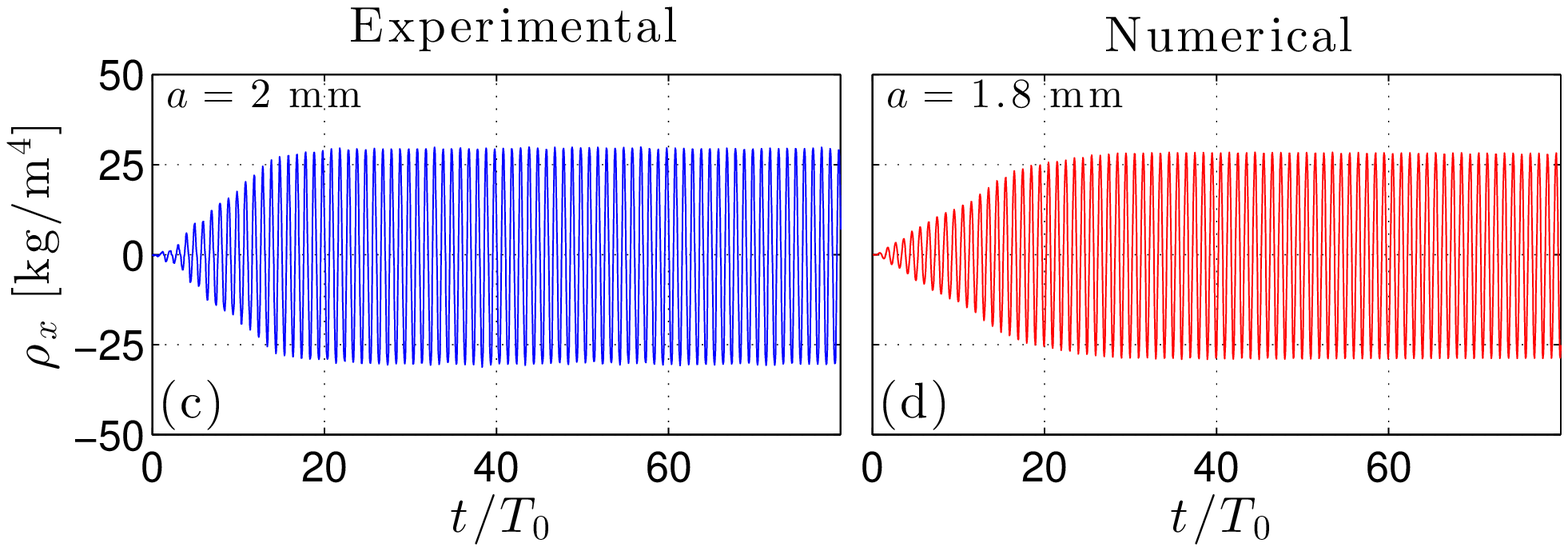}
  \vspace{-.3cm}
    \caption{Experimental (a) and numerical (b) snapshots of the horizontal density gradient at $t=50~T_0$. The amplitude of the wave maker is $a=2$~mm for the experiment and $a=1.8$~mm for the simulation. Both attractors are stable. Note that the shade (color online) scale is the same in both panels. The small black quadrilateral defines the acquisition region used for computing the time-frequency spectrum presented in figure~\ref{TimeFrequency}. The wave frequency is $\omega_0/N=0.62\pm 0.01$. Experimental (c) and numerical (d) horizontal density gradients as a function of the time, for a point located on the most energetic branch within the black trapezoid depicted in panels (a) and~(b). In the calculation we take a piecewise linear approximation of the experimental density profile, with the lower layer of depth $H'=308$ \ADD{mm} and buoyancy frequency $N$, and the upper layer of depth $\delta=18$ \ADD{mm} 
  with a density gradient 8 times smaller.     The total depth of fluid $H=H'+\delta=326$ \ADD{mm}. 
   \label{ComparisonStableCase}
   }
\end{figure}

As already mentioned in the Introduction, the dynamics of stable attractors has been studied in great detail before, both in stratified and rotating fluids. It is well established that the typical thickness of the attractor scales as $1/3$ power of the Ekman/inverse Stokes number in rotating/stratified fluid~\citep{RGV2001,Ogilvie2005,HBDM2008,GSP2008}. The evolution of the spatial wave spectrum during the transient regime after the start-up of the oscillations toward an equilibrium shape is also well-known~\citep{HBDM2008,GSP2008}. This dynamics is well reproduced in our calculations {be they 2D or} 3D. In addition, we observe a very good quantitative agreement in our 3D simulations once we introduce a small correction for non-perfect efficiency of the wave generator.

Typical snapshots of the computed and measured fields of the horizontal density gradient $\rho_{x}$ are shown in figure~\ref{ComparisonStableCase}, emphasizing a very good qualitative and quantitative agreement between the numerical and experimental data. 
This good agreement is further confirmed with the time-series of the horizontal density gradient at a point located in the most energetic branch of the attractor as illustrated in figure~\ref{ComparisonStableCase}. 
  Note that the numerical value of the density gradient is computed at the vertical mid-plane $xOz$ of the test tank.   With this comparison, we tacitly assume that the flow is approximately two-dimensional and the standard schlieren technique is applicable~\citep{DHS2000}. The effects of three-dimensionality are discussed further in the next section.

\subsection{An unstable attractor}

\begin{figure}
  \includegraphics[width=\linewidth,clip=]{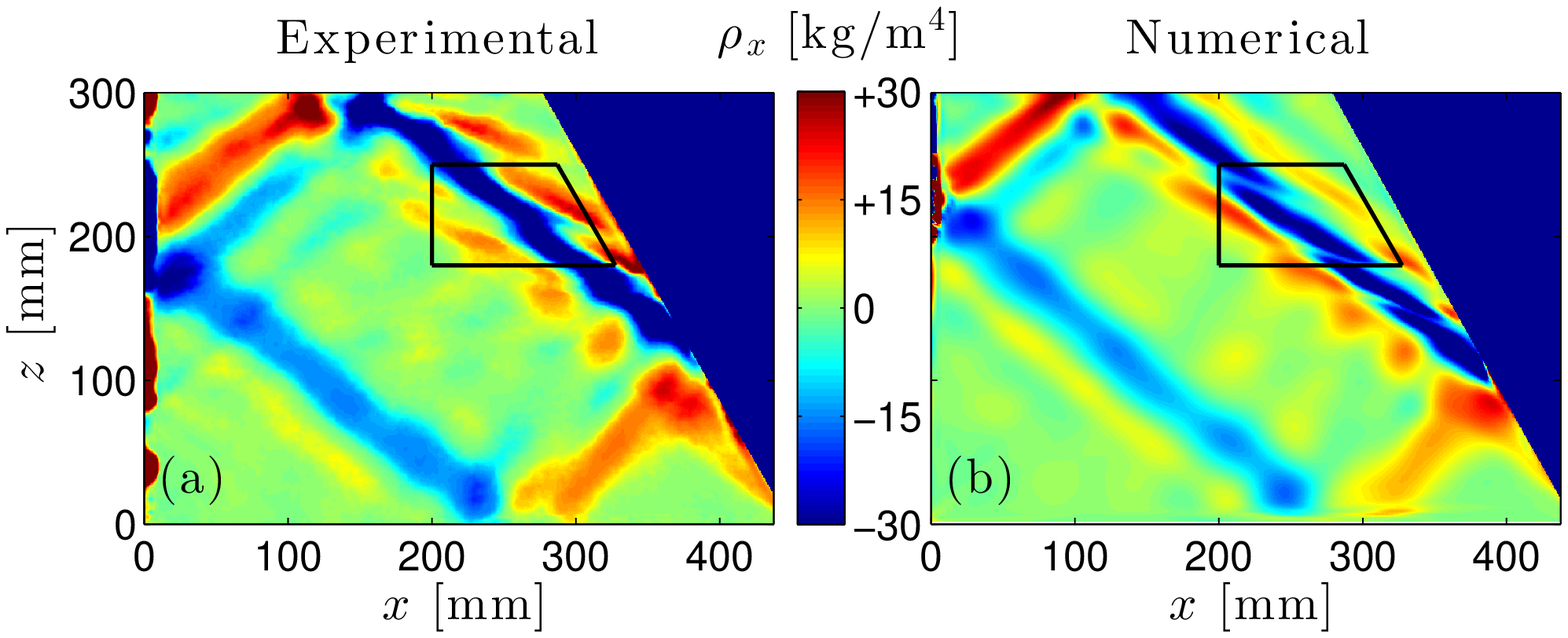}
  \vspace{-.3cm}
   \includegraphics[width=\linewidth,clip=]{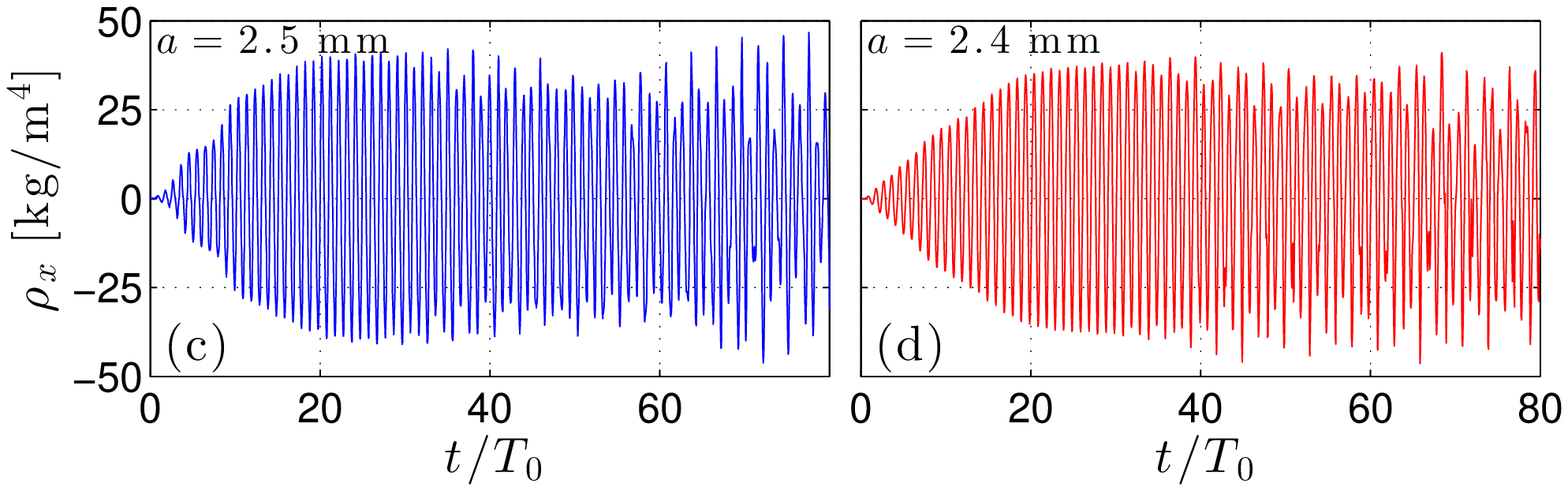}
  \vspace{-.3cm}
\caption{Experimental (a) and numerical (b) snapshots of the horizontal density gradient at $t=50~T_0$. The amplitude of the wave maker is $a=2.5$~mm for the experiment and $a=2.4$~mm for the numerical simulation. The wave frequency is $\omega_0/N=0.62\pm 0.01$. Both attractors are unstable and the instability appears first on the most energetic branch.  The small black quadrilateral defines the acquisition region used for computing the time-frequency spectrum presented in figure~\ref{TimeFrequency}. Experimental (c) and numerical (d) horizontal density gradients as a function of the time, for a point located on the most energetic branch within the black trapeze depicted in panels (a) and (b). The density stratification is the same as described in caption to figure~\ref{ComparisonStableCase}.
    \label{ComparisonUnstableCase}}
\end{figure}

As the amplitude of oscillations of the wave generator increases, the attractor becomes unstable. In what follows, we describe the comparison of the experimental
 and numerical results for the onset of the triadic resonance instability. 2D simulations were also performed and found to predict the instability to occur at the forcing amplitudes roughly two times smaller than the experimental ones. This is due to the absence of lateral walls in the 2D simulations. These walls introduce a significative dissipation in 3D simulations, as demonstrated in subsection~\ref{section:dissipation}. In view of this quantitative discrepancy between 2D and 3D simulations, in what follows we use the results of 3D calculations for comparison with the experiments. The experimental and numerical snapshots of the horizontal density gradient field are presented in figure~\ref{ComparisonUnstableCase}. The development of TRI is clearly seen in the most energetic branch of the attractor. A very good quantitative and qualitative agreement is again observed between the experimental and numerical wave fields. The experimental and numerical time-series at a point in the most energetic branch of the attractor  show also an excellent correspondence.

\begin{figure}
\begin{center}
  \includegraphics[width=0.85\linewidth,clip=]{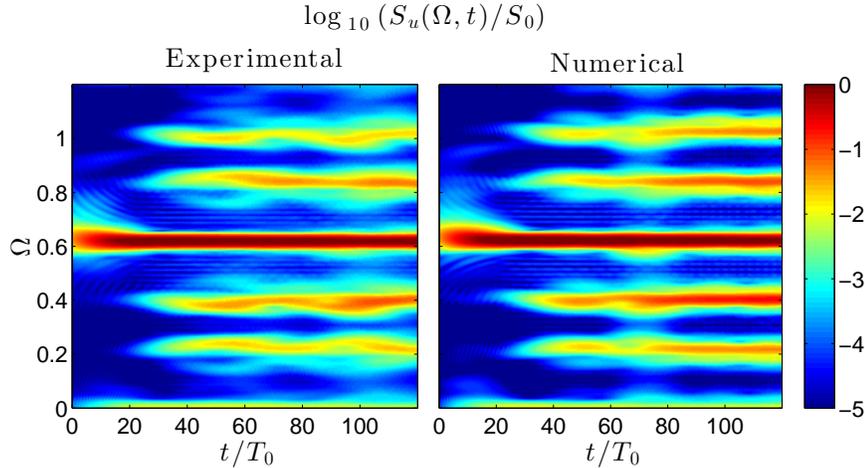}
    \caption{Experimental and numerical time-frequency spectra of data presented in figure~\ref{ComparisonUnstableCase}. Both have been computed with the same signal processing parameters, over the same area,
     located on the most energetic branch and depicted with a black trapeze in figure~\ref{ComparisonUnstableCase}.   \label{TimeFrequency}
   }
   \end{center}
\end{figure}

  Let us now focus on the development of the instability in more detail to see if computations reproduce the experimentally observed triads in temporal and spatial domains. 
 We first consider the evolution of the instability in the temporal domain. The development of the frequency spectrum of wave motion over time is presented in figure \ref{TimeFrequency}. The time-frequency diagrams are calculated from numerical and experimental data for points as in \cite{BDJO2013}, with the formula
\begin{equation}
S_r(\omega,t)=\left\langle \middle| \int_{-\infty}^{+\infty} \! v_r(x,z,\tau)e^{i\omega\tau}h(t-\tau)\, d\tau \middle|^2\right\rangle_{xz},
\end{equation}
where $h$ is an Hamming window and $r$ the component of the velocity field along $x$ or $z$, i.e. $S_u$ or $S_w$. 
The calculations are performed with the Matlab toolbox described in \cite{Flandrin1999}. The appropriate choice of the length of the Hamming window allows to tune the resolution in space and time. The experimental and numerical spectra are averaged over the analyzing area shown in figure~\ref{ComparisonUnstableCase}. It can be seen that the numerical and experimental spectra agree qualitatively and quantitatively.
The signal is initially entirely dominated by the forcing frequency $\Omega_{0}=0.62$
(in which $\Omega_{0}=\omega_{0}/N$ is the non-dimensional frequency of oscillation)
: it corresponds precisely to the primary (carrier) wave. 
Then oscillations with frequencies $\Omega_{1}$ and $\Omega_{2}$, which correspond to two secondary waves generated by TRI, slowly develop with time. At $t=50T_{0}$, these frequencies are $\Omega_{1}=0.24$ and $\Omega_{2}=0.38$. They satisfy the frequency conditions for the triadic resonance
\begin{equation}
\Omega_{1}+\Omega_{2}=\Omega_{0}.
\label{eq:PSI}
\end{equation}
In addition, one can see also two peaks $\Omega_{3}=0.86$ and $\Omega_{4}=1.00$ satisfying differential conditions
\begin{equation}
\Omega_{3}-\Omega_{1}=\Omega_{0} \qquad\mbox{and}\qquad
\Omega_{4}-\Omega_{2}=\Omega_{0}.
\end{equation}

To verify the fulfillment of the condition for triadic resonance in space, we apply the Hilbert transform technique~\citep{MGD2008,BDJO2013} to the numerically simulated data in the same way as it is done in experiments~\citep{BDJO2013}: the experimental signal is demodulated with the Fourier transform, filtered around the three frequencies of interest, $\Omega_{0}$, $\Omega_{1}$ and $\Omega_{2}$, and reconstructed back in real space using the inverse Fourier transform.  As shown in Figure~5 of \cite{SED2013}, it is possible to compute amplitude and phase of each component. The latter appear as patterns of stripes, corresponding to a fixed moment of time. The wave vectors can be derived by differentiating these phases along the $x$ and $z$ directions.

To quantify the wave vectors involved into the triadic resonance  in the numerical data, we construct probability density functions (PDF) for the components of wave vectors. Using these PDF for the components of the wave vector corresponding to the primary wave oscillating at $\Omega_{0}$, 
 we can estimate $\overrightarrow k_0 =  \{-63.5, -80.1\}$ $\mathrm{m}^{-1}$. 
  The same procedure for the secondary wave oscillating at $\Omega_{2}$ gives $\overrightarrow k_2 = \{  -104.8, -242.5  \}$ $\mathrm{m}^{-1}$. 
The phase pattern is slightly more complicated for the component oscillating at $\Omega_{1}$ as can be noticed in Figure~5 of \cite{SED2013} where lines of equal phase  for this frequency are not completely straight and there is a larger error bar on the measurement of the wave vector. 
 Numerically, it is possible to investigate further in the focusing branch of the attractor thanks to the PDF analysis: one realizes that there are two families of collinear wave vectors, which yield multi-peaked PDF of the wave vector components, where the main peak 
 corresponds to $\overrightarrow k_{1}=\{46.5, 174.3\}$ {$\mathrm{m}^{-1}$}. These values of the wave vector components closely match the spatial condition of the triadic resonance. The physical interpretation of the other peaks present in the PDF of the wave vector components for the wave field filtered at $\Omega_{1}$ is unclear.

\begin{figure}
 \begin{center} \includegraphics[width=0.8\linewidth,clip=]{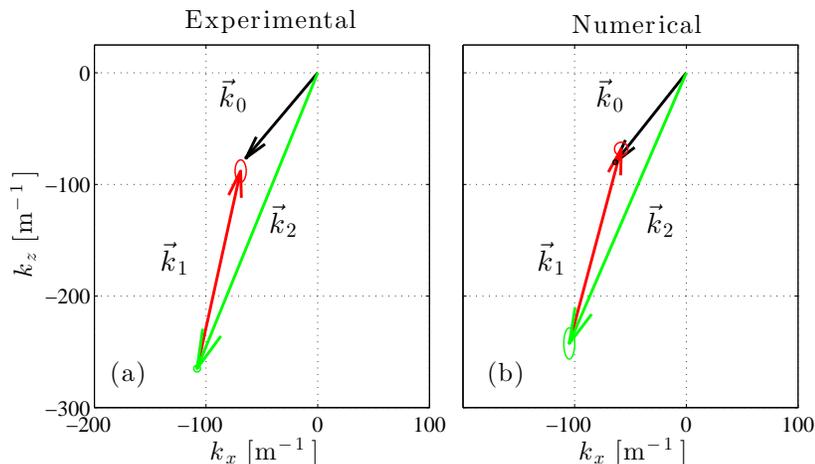}\end{center} 
  \vspace{-.3cm}
   \caption{Experimental (a) and numerical (b) triads.  The three wave vectors are measured from the phase of the waves after Hilbert filtering around $\omega_0$, $\omega_1$ and $\omega_2$. The error bars are given by of ellipses at the arrow tips.}
   \label{Triad}
\end{figure}
  The triangles of the wave vectors obtained in experiments and simulations are shown in figure~\ref{Triad}. Again, we observe a good quantitative agreement between the experimental and numerical results.

\section{Wave structure in the transversal direction: three-dimensional effects and role of lateral walls}
\subsection{Introduction}
 
\begin{figure}
\begin{center}
  \includegraphics[width=0.6\linewidth,clip=]{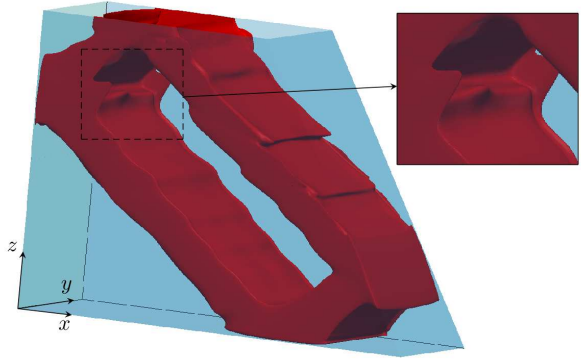}
\end{center}
  \vspace{-.3cm}
   \caption{Visualization of the three dimensional effects {in the internal wave attractor}. Snapshot of the instantaneous magnitude of the velocity field $(u^2+w^2)^{1/2}$  produced by the 3D numerical simulations based on the spectral element method for $a=2.4$~mm. The snapshot corresponds to a contour plot (level 2.5 \ADD{mm}/s) of the amplitude of the velocity field $(u^2+w^2)^{1/2}$ at $t=50~T_0$.The inset presents a zoom to emphasize an example in which the transversal direction is curved.
}   \label{ThreedimensionalEffect}
\end{figure}

We now consider the wave attractor in the whole domain and investigate its structure in the transversal direction. 
An overall idea of the importance of three-dimensional effects in wave attractors can be {indeed} drawn from Figure~\ref{ThreedimensionalEffect} which represents an isosurface  of the norm of the velocity vector for a 3D simulation at the onset of the instability. We can see the same features as in the vertical $xOz$ plane view in 
figure \ref{ComparisonUnstableCase} with the loop of the attractor and the secondary perturbations. However, the iso-surface of the velocity norm is not flat in the transversal direction as the fronts of the perturbations are visibly curved {especially on the two top S-shape features in the focusing branch near the slope.} 
Also, close to lateral walls, in particular in corner regions at the intersection of vertical walls{, it is possible to see some three-dimensional structures. The inset of figure~\ref{ThreedimensionalEffect} is a zoom on a region where the 3D-effects are particularly visible.} 

\subsection{2D visualizations in vertical plane $xz$ at different transversal locations}

In \cite{SED2013}, the experiments presented have been performed using the Synthetic Schlieren technique which assumes that the flow is bidimensional. Indeed,  in this technique,  the image results from the integration of the field along the transverse direction and the ray of light is expected to cross the tank through always the same density perturbation.  This assumption seems to be reasonable because the experimental set-up has been designed  on purpose   to ensure the two-dimensionality of the flow.    {Indeed,} the tank is narrow in the transverse $y$ direction and forcing is applied via the wave-maker through almost all the width of the tank so that the flow is expected to be two-dimensional, except very close to both vertical lateral walls where the viscosity plays an important role.

To check this assumption, we use experimental standard PIV technique for several vertical sections at different constant $y$ positions of the light sheet  and thus compare flow properties between the different sections. 
Experiments have been performed with $L$=435~\ADD{mm}, $H$=300 \ADD{mm} and $\alpha=25^\circ$,
 a geometry close to experiments performed in \cite{SED2013}. Different vertical sections in the transverse direction have been illuminated with a vertical laser sheet, coming from the bottom of the tank, which is transparent. A $45^{\circ}$ inclined mirror, located below and as long as the tank, transforms a horizontal laser sheet into a vertical one.  By simply translating the mirror in the transverse direction, one can illuminate the different test sections.  Three series of experiments have been performed, each one with a different amplitude of the wave maker: $a=1.5$, $3$ and $5$~mm. For each series, between $9$ to $12$ vertical sections 
  have been illuminated by simply translating the mirror in the transverse direction.

Measurements have been carried out in the stationary state of the attractor. For the lowest amplitude experiment, there is no instability of the attractor: the wave maker was  stopped only at the end of the full series, once all sections have been illuminated. For the two other series, the instability takes place after a long transient: consequently, the wave maker has been stopped after performing measurements at each transversal location 
 to avoid mixing effects that would modify the stratification and thus  the attractor itself. For all series, measurements in the central section of the tank were performed twice, once at the beginning, once at the end. This gives an estimate of the errors made on these measurements and also gives the possibility to check that, after a whole series, the attractor is really unchanged.
 To compare the different sections of the same series, we focused only on the most energetic branch of the attractor (the one
connecting the slope to the free surface). Transversal profiles are extracted through this branch, always at the same location. As the horizontal and vertical velocities are measured, it is possible to calculate the velocity along this branch inclined at the angle $\theta=\arcsin\Omega_0$ using the formula  
 $v_{\xi}=-v_x\cos\theta+v_z\sin\theta
$. 
 The maximum value corresponds to the center of the branch,  $\eta=0$, where $\eta$ is the coordinate transversal to the branch.
 
 Numerically, with 3D simulation, it is possible to get velocity profiles in the transverse $y$ direction. Two 3D simulations have been performed, with exactly the same geometrical parameters as in \cite{SED2013}, and with two different amplitudes of the wave maker: $a=1.6$ and $2.2$ mm. Velocity profiles in the transverse $y$ direction were extracted from the most energetic branch and the velocity $v_{\xi}$ along this branch computed as for experiments.
 
 It is important to note that the transverse resolution is much higher in the numerical data ($512$ points for $170$~\ADD{mm}) than in the PIV experimental ones (between $9$ and $12$ points for the same $170$~\ADD{mm}). Moreover, the different points obtained experimentally come from different individual experiments performed at the same input parameters but not from the same unique experiment. Thus, it is difficult to synchronize the different points and we are limited to look at amplitudes of the filtered signals at $\omega_0$, $\omega_1$ and $\omega_2$. 
In contrast, various quantities can be examined numerically such as the raw signal, the whole signal filtered around $\omega_0$, $\omega_1$ and $\omega_2$ and the mean flow generated by the attractor. Different videos, available in the supplementary information, show these different fields. 

 \subsection{Velocity profiles in transverse $y$ direction}\label{profiles}

 \begin{figure}
  \begin{centering}
   \includegraphics[width=\linewidth,clip=]{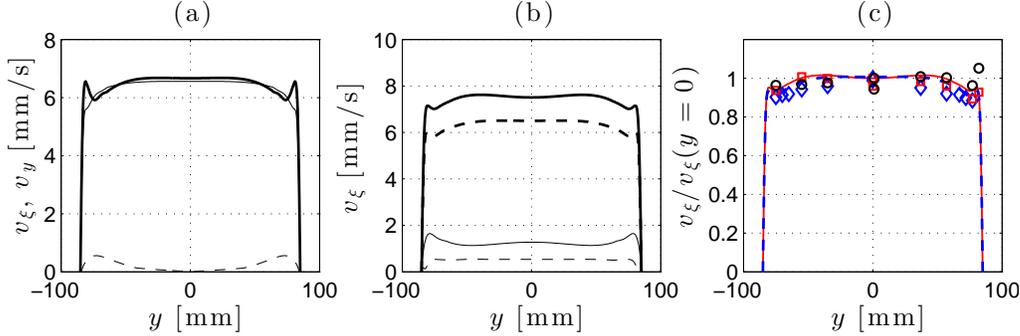}
   \caption{(a): Raw velocity profiles as a function of the transversal direction $y$ for the numerical simulation with $a=1.6$~mm. Thick solid line and thin solid line: maximum ($\max{v_{\xi}}$) and opposite of the minimum (-$\min{v_{\xi}}$) of the velocity $v_{\xi}$ for each $y$ over the time history. Dashed line: maximum of the absolute value of the transversal velocity $v_y$ over the time history for each~$y$. (b):~Modulus of $v_{\xi}$ filtered over $20~T_0$ around $\omega_0$ (thick dashed line), for the numerical simulation with $a=1.6$~mm. Modulus of $v_{\xi}$ filtered over $20~T_0$ around $\omega_0$ (thick solid line), $\omega_1$ (thin solid line) and $\omega_2$ (thin dashed line), for the numerical simulation with $a=2.2$~mm. (c): Normalized profiles of $v_{\xi}$ filtered around $\omega_0$ for the numerical simulation with $a=1.6$~mm (thick dashed blue line) and $a=2.2$~mm (thin solid red line). Points were obtained using PIV experiments: $a=1.5$~mm (blue diamond), $a=3$~mm (red square) and $a=5$~mm (black circle). }
   \label{TransversalCut}
     \end{centering}
\end{figure}

 Figure~\ref{TransversalCut}(a) presents the different velocity profiles for the numerical simulation with $a=1.6$~mm. For this amplitude, the attractor is stable. The raw velocity profiles for $v_{\xi}$  are, as expected, very close to the bidimensionality, except into the boundary layers where the velocity goes to zero. Moreover, it appears clearly that the amplitude of the transverse velocity $v_y$ (represented by a dashed line) is much smaller than the velocity $v_{\xi}$ along the branch. This is a clear confirmation that the attractor generated by the wave maker is two-dimensional to a good approximation. The difference between the raw velocity magnitudes measured in positive and negative $\xi$ directions comes from the presence of slow near-wall secondary currents superposed with the fundamental monochromatic wave motion in the wave beams as discussed in more detail in \S 4.4.
 
To extract only the component of the wave field oscillating at the fundamental frequency, one can do Hilbert filtering on the raw profiles around $\omega_0$ over $20~T_0$. The modulus of the velocity filtered for the numerical simulation with $a=1.6$~mm is shown with a thick dashed line in figure~\ref{TransversalCut}(b). As the raw profiles, it is also very close to bidimensionality. 
By looking to the real part of the filtered signal, 
one can be convinced that the profiles for $v_{\xi}$ are exactly the same for positive and negative values.
 Hilbert transform allows us also to investigate the two-dimensionality of the TRI. As the numerical simulation with $a=1.6$~mm is stable, one can use the one with $a=2.2$~mm, which is unstable. Figure~\ref{TransversalCut}(b) shows the profiles along $y$ of the modules of $v_{\xi}$ filtered around $\omega_0$ (thick solid line), 
$\omega_1$ (thin solid line) and $\omega_2$ (thin dashed line) for this simulation. These profiles appear to be nearly constant in the $y$ direction, except close to the edges. This ensures that the instability which develops in the tank is also bidimensional.

To compare the numerics and the experiments, we can superimpose the experimental and numerical $v_{\xi}$ profiles after Hilbert filtering around $\omega_0$ for the different amplitudes of the wave maker. Figure~\ref{TransversalCut}(c) shows these different profiles: the solid blue line (resp. diamonds) corresponds to numerical (resp. experimental) data for $a=1.6$~mm (resp. $a=1.5$~mm) while the dashed red line (resp. squares) corresponds to numerical (resp. experimental) data for $a=2.2$~mm (resp. $a=3$~mm). Experimental points for $a=5$~mm are represented by black circles. All data are normalized by the velocity in $y=0$~\ADD{mm} to avoid divergences in amplitude due to the location of the point where is measured the velocity along the branch and small discrepancies in wave maker amplitude. All data show again a very good two-dimensionality. The loss of amplitude on the edges, before the boundary layers is less than $10\%$. {Moreover, numerical and experimental data for $a=1.6$ and $a=1.5$~mm (dashed blue line and diamonds) show a good agreement between each other. For higher amplitudes $a=2.2$ and $a=3$~mm (red line and squares), where the TRI is developed, the agreement is also very good. Even for the experiment with $a=5$~mm, where TRI is well developed, the shape of the experimentally measured transverse profile is in reasonable agreement with the simulation at $a=2.2$ mm. 
	All numerical and experimental data collapsed in a reasonable way on a curve that ensures that all waves inside the tank are reasonably invariant in the transverse $y$-direction.}

\subsection{Generation of mean-flows}\label{meanflow}

We previously showed that the velocity field in the transverse direction for the wave attractor confirms the usual 2D approximation as a reasonable one. Nevertheless, it turns out that a closer look on the profiles of $v_{\xi}$ reveals a discrepancy between profiles of the maximum (max $v_{\xi}$) and the opposite of the minimum (-min $v_{\xi}$) of the velocity~$v_{\xi}$ as illustrated in figure~\ref{TransversalCut}(a). Indeed, the raw profiles for the attractor are not totally symmetric: when $v_{\xi}$ is positive (thick solid line), even if the profile looks two-dimensional, there are some small peaks close to boundary layers which are not present in the profiles when $v_{\xi}$ is negative (thin solid line). This symmetry breaking suggests the presence of a mean flow, always in the same direction, close to the boundary layers and which modifies the velocity profiles close to boundary layers.
To verify this, we analyse now further the data and check for possible mean-flow generated within the tank.
To extract this mean-flow from the raw data, we used Hilbert filtering around $\omega=0$~rad/s.

The numerical simulation {specially run to thoroughly study} the mean-flow has been performed with the same geometrical parameters as the one with the 
 amplitude $a=1.6$~mm for section~\ref{profiles}. The attractor is stable and is precisely located at the same position in the tank as in figure~\ref{ComparisonStableCase}. To limit the size of the data saved, the three components of the velocities were recorded  only for the two horizontal planes, $z=100$~\ADD{mm} and $z=200$~\ADD{mm}, with $0.5$~second time-steps. On theses planes, the velocities were saved on a mesh with a good resolution of $128 \times 256$ points.

\begin{figure}
   \begin{centering}
        \includegraphics[width=0.85\textwidth]{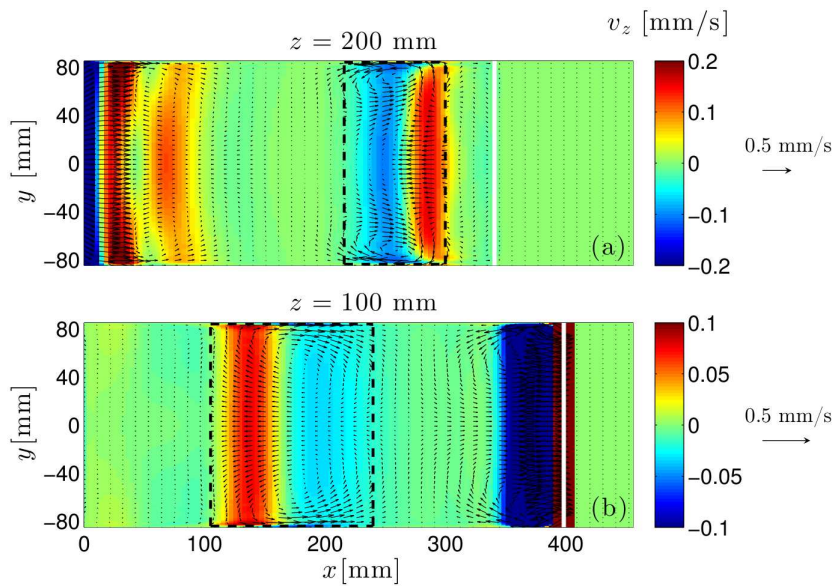}
   \caption{Mean-flow (Hilbert filtering analysis  at $\omega=0$~rad/s applied on numerical data) generated in the two different horizontal planes  $z=200$~\ADD{mm} (panel (a)) and $z=100$~\ADD{mm} (panel (b)). The color represents the vertical velocity component, $v_z$, while arrows represent the horizontal velocity components, $v_x\, \vec{x}+v_y\, \vec{y}$, with the arrow scale displayed at the right of the figure. 
   The wave maker is located at $x=0$~\ADD{mm}, on the left of the figure, while the slope is indicated, on the right, by a vertical white line. 
The dashed rectangles delimit the regions where the attractor beams intersect with the planes located at $z=200$~\ADD{mm} (panel a)  and  $z=100$~\ADD{mm} (panel b). 
}
\label{fig:meanflow}
\end{centering}
\end{figure}

Hilbert filtering at $\omega=0$~rad/s was applied on {time-series} of the three velocity components for each point of the $x$-$y$ plane. We consider the situation once the attractor reaches the steady state.
 Figure~\ref{fig:meanflow} shows the mean-flow in the two different horizontal planes,  $z=100$~\ADD{mm} and $z=200$~\ADD{mm}. The color represents the vertical velocity component~$v_z$, while arrows represent the horizontal velocity components, $v_x\, \vec{x}+v_y\, \vec{y}$. The wave maker is located at $x=0$~\ADD{mm}, on the left of the figure. The slope is indicated, on the right, by a vertical white line. Figure~\ref{ComparisonStableCase} shows the location of the attractor in the $x-z$~plane. At $z=100$~\ADD{mm}, the attractor is reflecting on the slope while, at $z=200$~\ADD{mm}, it is reflecting on the wave maker. These are the two intense regions in figure~\ref{fig:meanflow}, close to the slope at $z=100$~\ADD{mm} and close to the wave-maker at $z=200$~\ADD{mm}. As reflections occur in these regions, it can be difficult to extract conclusions for the mean-flow.
Nevertheless, it is still possible to examine the mean-flow in the regions where the attractor beams intersect with the planes located at $z=200$~\ADD{mm} and  $z=100$~\ADD{mm}. These intersection regions are delimited by the dashed rectangles in figure~\ref{fig:meanflow}. The flow in the branches of attractors shows a clear tridimensional behavior, with recirculation zones where the horizontal components of the velocity vectors are forming jet-like currents close to the vertical walls of the tank to compensate the mean flow in the central part of the wave beam.
It is important to emphasize that the magnitude of the mean-flow is much smaller than the velocity of the fluid due to the waves. Indeed, for this attractor, the typical amplitude of the velocity oscillations in the most energetic branch at the intersection with the plane  $z=200$ \ADD{mm} is about $2.5$~\ADD{mm}/s while the typical velocities of the mean-flow are about  $0.1$~\ADD{mm}/s, thus less than $5\%$ of the velocity due to the waves.

\subsection{Dissipation in the bulk and in the boundary layers\label{section:dissipation}}

\begin{figure}
   \begin{centering}
        \includegraphics[width=0.85\textwidth]{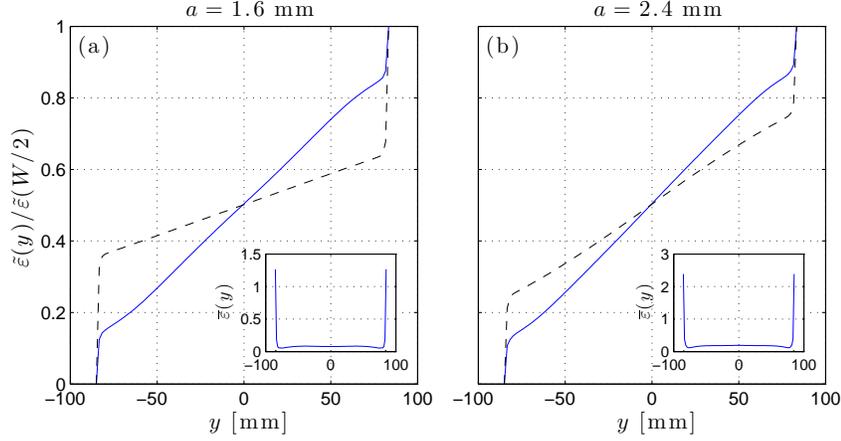}
   \caption{Horizontally integrated dissipation $\tilde \varepsilon (y)$
    (see definition in Eq.~\ref{defdissipationintegree}), normalized by the total dissipation $\tilde \varepsilon (W/2)$.
Panel (a) corresponds to the forcing amplitude $a=1.6$~mm, while panel (b) to $a=2.4$~mm. 
The solid blue line corresponds to a region located within the most energetic branch of the attractor,
while the dashed black line corresponds to a region close to the center of the tank 
outside the attractor. The inset present the locally averaged dissipation $\overline \varepsilon(y)$ (in \ADD{mm}$^2\cdot$s$^{-3}$) for both forcing amplitudes within the most energetic branch.}
\label{fig:dissipation}
\end{centering}
\end{figure}

To further examine the three dimensionality of the flow, we determine the spatial distribution
of the dissipation in the tank.
We first compute the  dissipation 
 \begin{equation}
\varepsilon(x,y,z,t)=2\nu e_{i,j}e_{i,j}\label{defdissipation}
 \end{equation}
 where $\nu$ is the kinematic viscosity while the strain rate tensor is defined as $e_{i,j}=\frac{1}{2}\left(\frac{\partial v_i}{\partial x_j}+\frac{\partial v_j}{\partial x_i}\right)$ where $v_i$ and $x_i$ are respectively the components of the velocity field and of the position.
 Then, for a given $y$-transversal coordinate, we average on a 
$20\times20$ \ADD{mm}$^2$ square to define the locally averaged dissipation
 \begin{equation}
\overline \varepsilon(y)=\langle\, \varepsilon(x,y,z,t)\, \rangle _{(x,z,t)},\label{defdissipation}
 \end{equation}
 which has also been averaged, after 42 periods after the start of the forcing,  over three time periods.
 This quantity is shown as an inset in figure~\ref{fig:dissipation}.
 
However, as for example in \cite{BBM2008}, 
 it is even more explicit
 to integrate this result horizontally along the transversal $y$-coordinate  
  as follows
 \begin{equation}
\tilde \varepsilon (y)=
{\displaystyle \int_{-W/2}^y \overline \varepsilon(y')\ \mbox{d} y'}
\label{defdissipationintegree}
\end{equation}
in which $W$ is the total width of the tank.

We plot in figure~\ref{fig:dissipation} the horizontally
integrated dissipation $\tilde \varepsilon (y)$, normalized by the total dissipation $\tilde \varepsilon (W/2)$
within the parallelepiped 20 \ADD{mm}$\times20$ \ADD{mm}$\times W$,
in two different locations: one in the most energetic branch
after the focusing onto the slope (with the subscript $a$ for attractor), the second one around the center of the tank (with the subscript $c$ for center) and thus far from any
branch of the attractor. For a weak forcing
$a$=1.6 mm which leads to a stable attractor, one gets
$\tilde \varepsilon_a (W/2)=  \ADD{1.7}$ and 
$\tilde \varepsilon_c (W/2)=0.013$~\ADD{mm}$^2\cdot$s$^{-3}$.
For a larger forcing $a=2.4$ mm leading to unstable regime, above values are typically multiplied by a factor 2, since one gets
$\tilde \varepsilon_a (W/2)=\ADD{3.7}$ and
$\tilde \varepsilon_c (W/2)=0.028$~\ADD{mm}$^2\cdot$s$^{-3}$.

Above figure~\ref{fig:dissipation} emphasizes that, within the most energetic branch, the spatial distribution 
of the dissipation is only weakly changed while increasing the amplitude of the forcing, and therefore 
by passing from a stable to an unstable attractor. Approximately 1/4 of the dissipation 
is located in the boundary layers, and the three remaining quarters are within the bulk of the flow.
The dissipation outside the attractor is significantly altered by the amplitude of the forcing. This is presumably
because TRI has generated secondary waves which propagate, outside the attractor
and more precisely toward the center of the tank.
However, it is important not to overestimate this result since the total dissipation in the 
center of the tank is more than 100 times smaller than its counterpart within the most energetic
branch of the attractor. It was of course expected, since this is where waves, and therefore energy,
 are trapped.
 Moreover, figure~\ref{fig:dissipation} reveals that, 
in the most energetic branch of the attractor, approximately 25\% of the total dissipation occurs within the boundary layers which occupy less than 10\% of the total width.

\subsection{Synthetic Schlieren test}

\begin{figure}
   \begin{centering}
        \includegraphics[width=0.8\textwidth]{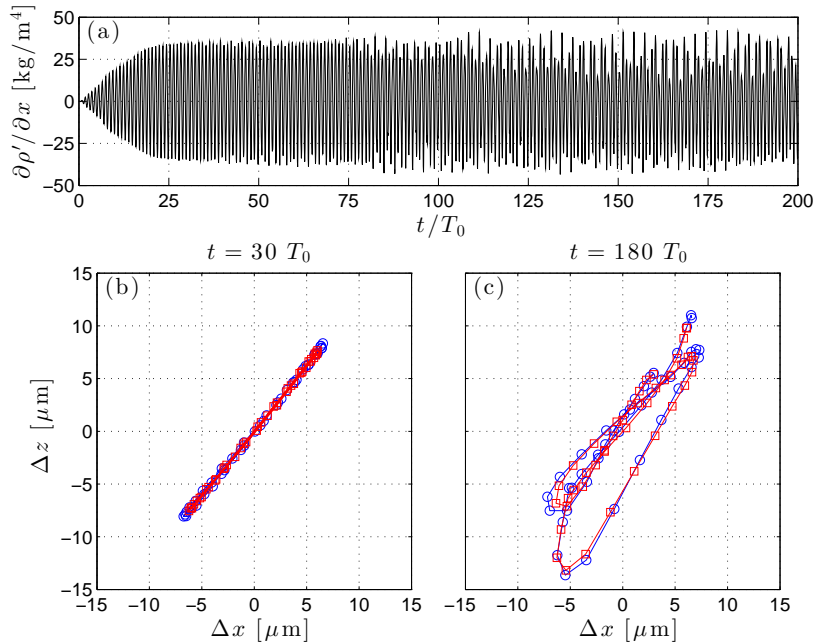}
   \caption{(a) Time history of horizontal density gradient in $y=0$~\ADD{mm} for the numerical simulation with $a=2.2$~mm.
   (b) Displacement of the dots on the CCD camera that the simulation would have caused if the density gradients have been measured using Synthetic Schlieren technique at $t=30~T_0$, before the onset of TRI. (c) same as panel b) after the onset of TRI ($t=180~T_0$). The blue circles indicated the displacement caused if the flow is assumed to be $2$D, while the red squares show the displacement caused without this assumption. The displacements are plotted for two periods of the wave-attractor around the time indicated above the figures.}
\label{fig:optic}
\end{centering}
\end{figure}

Using the numerically computed density gradients in the transversal direction, we can provide an estimation of the error made on experimentally measured density gradients by assuming that the flow is bidimensional inside the tank.
We use the simulation for $a=2.2$~mm as in section~\ref{profiles}, by considering the density gradients instead of velocity fields. These data were available on a point located on the most energetic branch, for all $y$. Figure~\ref{fig:optic}(a) shows the time history of the horizontal density gradient in $y=0$~\ADD{mm}. TRI occurs after $35~T_0$ in this simulation.

\cite{DHS2000} have shown that the angle of deflection of a light ray crossing the tank in the $i$-direction is 
\begin{equation}
\alpha_i=\frac{1}{n}\frac{\mbox{d} n}{\mbox{d} \rho}L\frac{\partial \rho'}{\partial i},
\label{alpha1}
\end{equation}
$i$ being $x$ or $z$. Here $L$ is the width of the tank and $n$ the optical index of water. The quantity ${\mbox{d} \rho}/{\mbox{d} n}$ is essentially constant and equal to $4.1\times10^3$~kg/m$^3$. Equation~(\ref{alpha1}) assumes that the flow is bidimensional and that the density gradients do not depend on the $y$-direction. This deflection induces displacement of the dots on the camera CCD from which we can measure the density gradients.

With the numerical simulations, one can proceed in the inverse way. Knowing the density gradients and especially their dependence in $y$, one can compute the displacement that these gradients would have caused on a CCD camera placed at typical experimental distances. For density gradients depending slightly on $y$, one has to integrate over the width of the tank to get the deflection
\begin{equation}
\alpha_i=\frac{1}{n}\frac{\mbox{d} n}{\mbox{d} \rho}\int_0^L\frac{\partial \rho'}{\partial i}\mbox{d}y.
\label{alpha2}
\end{equation}

This integration assumes that the deflection of the ray is small in $x$ and $z$ directions, so the ray remains more or less at $(x,z)$ constant. That is why one can use only a line profile in $y$ for density gradients.
In typical experimental conditions, the CCD camera is placed at $1.75$~m of the tank and at $2.25$~m of the random dot pattern. The camera has a lens of $25$~mm.  
Considering these experimental parameters, simple geometrical calculations then allows us to convert the angle deflection $\alpha_i$ into the displacement on the camera $\Delta_i$. Displacements are indicated
on figures~\ref{fig:optic}(b) and~(c) by the red squares, before the onset of TRI (panel b) and after the onset of TRI (panel c). The order of magnitude of these displacements is of few micrometers, while the pixel size of the two different cameras used are equal to $3.45$ and $4.65$~$\mu$m. Thus, the displacements are of few pixels and sufficient for data treatment. Interestingly, 
one can also compute the displacements as if the flow was strictly bidimensional, by assuming that, once given $(x,z)$, the density gradients in $y=0$  are the same for all $y$ in the tank. The deflection is thus as in equation~(\ref{alpha1}). These displacements are indicated on figure~\ref{fig:optic}(b) and (c) by the blue circles, before the onset of TRI (panel b) and after the onset of TRI (panel~c). The difference between  blue circles and red squares is very small in both situations, before and after the onset of TRI. This shows that the bidimensional assumption for the flow is robust and the error using it is very small. One can see on figure~\ref{fig:optic}(b) and (c) that the displacements of blue circles are a slightly larger than those for red squares. This shows that the density gradients in the center plane $xOz$  are slightly underestimated in the experimental measurements performed with the synthetic schlieren technique, typically by a factor of $5\%$ without TRI and by a factor of $7\%$ with TRI. This combined experimental-numerical check is a nice confirmation of the validity of the synthetic schlieren technique that justifies the comparisons performed in \S 3.1 and \S 3.2 of the present paper.

\section{Conclusion}

In this paper, we present the comparison of the experimental and numerically simulated internal wave attractors in a trapezoidal domain filled with a linearly stratified fluid. Three-dimensional versions of the spectral element code Nek5000~\citep{FischerRonquist1994} have been used in the numerical simulations. 
As expected, in the linear regime, we recover the dynamics of the wave attractors, which is well described in the literature~\citep{HBDM2008,GSP2008,HGD2011}. 

In the nonlinear regime, three-dimensional numerical simulations correctly reproduce the dynamics of the experimentally observed triadic resonance~\citep{SED2013} in terms of spatial and temporal parameters of primary and secondary waves (frequencies, wave numbers).
 The measurements of the fine parameters of the calculated internal wave patterns are performed by applying exactly the same technique as in experiments~\citep{SED2013}, with a systematic use of the Hilbert transform~\citep{MGD2008,BDJO2013}. This yields very strong quantitative evidence that, both numerically and experimentally, we observe exactly the same scenario of triadic resonance. Recent two-dimensional calculations of the nonlinear instability in rotating fluids~\citep{JouveOgilvie2014} are qualitatively similar but less conclusive quantitatively due to the lack of comparison with the experimental data.

Three-dimensional numerical simulations give access to distribution of wave amplitudes across the test tank and the mean wave-induced currents. The numerically simulated distribution of wave amplitudes across the tank is in good agreement with the experimental data. The transversal distribution of the velocity field reveals interesting features: it typically has secondary maxima close to lateral walls of the test tank. These maxima are due to a mean-flow generated by the wave-attractor. The mean-flow is strong close to the boundary of the tank but the magnitude of the mean-flow velocity remains small in comparison with the amplitude of velocity oscillation in the wave beams. Moreover, in the most energetic branch of the attractor, we have shown that approximately 25\% of the total dissipation occurs within the boundary layers which occupy less than 10\% of the total width. Three-dimensional calculations allow us to estimate the error introduced in conventional synthetic schlieren technique which assumes two-dimensionality of the flow and considers spanwise integrated optical distorsions. For numerically generated density fields closely reproducing the experiment ones, the error is found to be small, typically about 5\%, thus validating the usual bidimensional assumption used in such experiments. 
The validation of three-dimensional spectral element code in a present nominally two-dimensional problem represents a necessary step for future exploration of fully 3D problems in the spirit of~\cite{MandersMaas2004}, \cite{Maas2005} and \cite{DM2007}, which can shed light on the occurrence of attractors in oceanographic problems. Another line of attack for the future research is the exploration of long-term behaviour of internal wave attractors generating turbulence-like regimes, which have many similarities with the regimes observed in rotating systems~\citep{Favier-etal-2015}. Finally, the question of energy budget in the system, in the spirit of~\cite{JouveOgilvie2014}, is to address. This has to be done with emphasis on quantification of the mixing efficiency in a wide range of forcing (linear and nonlinear regimes) and in wide ranges of Reynolds and Schmidt numbers. The question of turbulent diffusivity, defined in~\cite{CXCPF2005}, is also very relevant.

\section*{Acknowledgements}
EVE gratefully acknowledges his appointment as a Marie Curie incoming fellow
at  ENS de Lyon. We thank P. Flandrin and S. Joubaud for helpful discussions.
 This work has been partially supported
by  ONLITUR grant (ANR-2011-BS04-006-01), by Russian ministry of education (RFMEFI60714X0090), RFBR (15-01-06363) and CFD web-laboratory unihub.ru. It has  been
achieved thanks to the resources of PSMN from ENS de Lyon. Most of numerical simulations were performed on supercomputer ``Lomonosov" of Moscow State University.

\bibliographystyle{jfm}

\end{document}